\documentclass[conference]{IEEEtran}
\IEEEoverridecommandlockouts
\usepackage{amsmath,amssymb,amsfonts}
\usepackage{algorithmic}
\usepackage{graphicx}
\usepackage{textcomp}
\usepackage{xcolor}
\usepackage[numbers,sort&compress]{natbib}
\usepackage{multirow}

\def\BibTeX{{\rm B\kern-.05em{\sc i\kern-.025em b}\kern-.08em
    T\kern-.1667em\lower.7ex\hbox{E}\kern-.125emX}}
\begin{document}

% \title{Unsupervised Learning of Image Distortion Maps}
\title{JOINT DEEP IMAGE RESTORATION AND UNSUPERVISED QUALITY ASSESSMENT}

\author{\IEEEauthorblockN{Hakan Emre Gedik, Abhinau K. Venkataramanan, and Alan C. Bovik}
\IEEEauthorblockA{\textit{The University of Texas at Austin, Austin, TX 78705 USA} \\
Email: \{hakan.gedik, abhinaukumar\}@utexas.edu, bovik@ece.utexas.edu}
}

\maketitle

\begin{abstract}
Deep learning techniques have revolutionized the fields of image restoration and image quality assessment in recent years. While image restoration methods typically utilize synthetically distorted training data for training, deep quality assessment models often require expensive labeled subjective data. However, recent studies have shown that activations of deep neural networks trained for visual modeling tasks can also be used for perceptual quality assessment of images. Following this intuition, we propose a novel attention-based convolutional neural network capable of simultaneously performing both image restoration and quality assessment. We achieve this by training a JPEG deblocking network augmented with ``quality attention'' maps and demonstrating state-of-the-art deblocking accuracy, achieving a high correlation of predicted quality with human opinion scores.
\end{abstract}


\section{Introduction}
Digital images suffer from a wide variety of distortions during their capture, compression, and transmission. Although the best way to assess perceptual image quality is to conduct large-scale subjective human studies, it is usually not feasible since they are burdensome and costly. Image quality assessment (IQA) is mainly concerned with designing algorithms that predict the likely human opinion score of an image. IQA methods are mainly classified into three categories based on the information available about the pristine reference image. If the pristine source image is available along with the distorted image, Full Reference (FR) IQA methods~~\cite{ssim, GMSD} can be employed. When the reference image is not available but partial information about it is at hand, Reduced Reference (RR) IQA methods~\cite{strred, speed} are used. No Reference (NR) methods~~\cite{brisque, niqe} address scenarios in which there is no reference information. Although FR methods achieve high correlations against subjective opinion scores, in real-world scenarios, the pristine reference image may not be available. This is particularly the case when ingesting user-generated content, which may already be distorted (e.g. compressed). Therefore, NR quality models are key tools in processing such images.    

Having showcased their ability to construct robust visual representations in various machine vision tasks, deep learning methods have also found growing adoption in the field of image quality assessment. Kang et al.~\cite{nrcnn} developed a deep network that consists of a convolutional layer with max and min pooling and two fully connected layers for NR IQA. In~\cite{paq2piq}, Ying et al. introduced PaQ-2-PiQ, a deep NR IQA model that incorporates a region proposal network~\cite{fastrcnn, fasterrcnn} to improve global quality predictions using estimated local quality.

At the same time, hidden activations of deep neural networks trained for machine vision tasks, such as image recognition, have also been shown to be perceptual quality-aware. Li et al.~\cite{perceptualLi} proposed a perceptual training loss function for general image transformation tasks based on the activations of a pre-trained VGG-16 network~\cite{vgg-16}. This loss function may be adopted as a perceptual FR IQA metric. This idea was then generalized to other deep architectures by~\cite{unreasonableperceptual}. IQA methods based on the distance of deep features surpass traditional methods in some cases, especially when assessing geometric distortions~\cite{unreasonableperceptual}, owing to the sensitivity of deep features to high-level semantic information. Interestingly, they have yet to be successfully deployed for assessing the quality of compressed media~\cite{funque_plus}.

The reverse has also been shown to be true. Quality-aware features have been used to improve image restoration performance by FBCNN~\cite{fbcnn}, which uses quality-aware training to perform deblocking on JPEG-compressed images over a wide range of compression levels. In FBCNN, the compression level of the input image is estimated, based on which ``quality factor'' embeddings are computed and used to modulate the image reconstruction process. By contrast, prior deep learning-based methods~\cite{arcnn, mwcnn} required training a separate model for each quality level.             

Although the quality factor embeddings in FBCNN~\cite{fbcnn} are explicit quality-related features, the information they carry about the distorted regions of the image is limited. Specifically, the quality factor embeddings of FBCNN are learned from a single estimated quality score, which discards most of the information about the spatial locations and severities of the distortions.

% PaQ-2-PiQ~\cite{paq2piq} generates spatial quality maps using the region proposal network. These maps function as masks, detecting high-quality patches of the image. However, they are not dependent on the type of distortion; instead, they only identify the salient regions of the image.

Therefore, all the aforementioned methods typically solve one task, either machine vision, image restoration, or quality assessment, and may use each other to improve their accuracy. Here, we formulate a deep neural network that is designed to simultaneously tackle both the image restoration and unsupervised quality assessment tasks. We achieve this through the use of ``residual quality attention'' (RQAttention), which is a modified residual attention~\cite{res_atten} mechanism. 

% with the primary objective of inferring image distortion maps. These distortion maps are intended to exhibit the following key properties:

% \begin{itemize}
%     \item Identification of distorted regions and quantification of distortion severities.
%     \item Classification of the type of distortion present in each spatial location.
%     \item Visual interpretability to enhance the comprehensibility of the generated maps.
% \end{itemize}

\begin{figure}[tb]
\begin{center}
\includegraphics[width=0.75\linewidth]{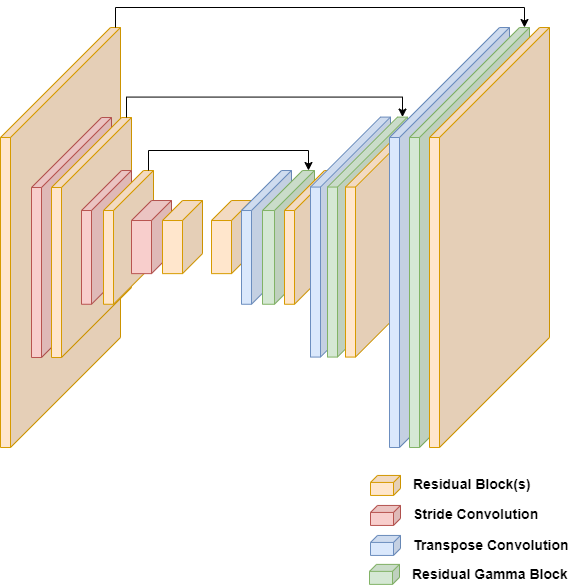}% This is a *.eps file
\end{center}
\caption{Model QAIRN Architecture}\label{fig:overall architecture}
\end{figure}

\section{Proposed Method}
\subsection{Network Architecture}
Inspired by the success of FBCNN~\cite{fbcnn}, we adopted a residual U-Net architecture \cite{resnet, unet} as the ``trunk'' encoder-decoder model that performs image restoration. This is similar to the architecture of FBCNN, with the Quality Factor Attention (QFAttention) blocks replaced by standard residual blocks. In addition, we introduce a novel attention block to our network called Residual Quality Attention (RQAttention).

\begin{figure}[tb]
\begin{center}
\includegraphics[width=\linewidth]{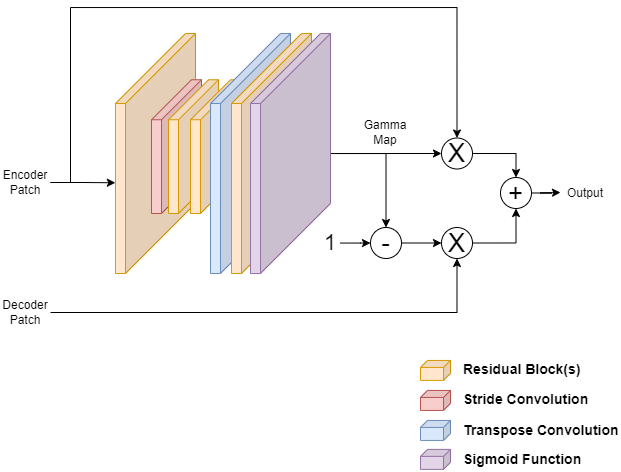}% This is a *.eps file
\end{center}
\caption{Residual Quality Attention Architecture}\label{fig:gamma block}
\end{figure}

RQAttention blocks, shown in Figure \ref{fig:gamma block}, consist of small autoencoders that predict a single-channel attention map \(\gamma_n(i,j) \in [0, 1]\). This is achieved using strided convolutions and transpose convolutions for down and upsampling, and a sigmoid activation at the output layer. The attention map is then used to gate all channels of the skip connection from the encoder and the output from the decoder. Specifically, the output of the RQAttention block is computed as
\begin{equation}
    \Tilde{x}_n = \gamma x_{skip, n} + (1 - \gamma) x_{decoder, n}
\end{equation}

The key insight is that for high-quality regions, the encoder features can bypass further processing using the skip connection, i.e. \(\gamma(i,j)\) is high, since they require minimal restoration. On the other hand, the worst quality regions in the input image require a greater deal of restoration, because of which the skip connection is suppressed and the trunk decoder's output is used, i.e., \(\gamma(i,j)\) is low. 

Therefore, the attention map implicitly encodes the quality of local regions, hence its interpretation as a \textit{quality} attention mechanism. Note that this property arises directly from the restoration and attention processes without requiring quality labels during training time. As we show in Section \ref{sec:evaluation}, quality attention maps are indeed highly correlated with subjective opinion scores of image quality.    

The overall model architecture including the U-Net trunk and RQAttention layers is illustrated in Fig. \ref{fig:overall architecture}. Due to the use of RQAttention, we name this model the Quality-Aware Image Restoration Network (QAIRN). 

% Each yellow block on the encoder side corresponds to 4 stacked residual convolutional blocks~\cite{resnet}, whereas on the decoder side, we use 6 of them. For downsampling and upsampling, we use strided convolutions and transpose convolutions with strides 2, respectively. Gamma blocks are introduced right after transpose convolutions on the decoder side.

% We name this residual convolutional autoencoder-based network with the introduced gamma blocks  

\subsection{Training Procedure}
To demonstrate the effectiveness of RQAttention, and to compare its restoration performance against that of FBCNN, QAIRN was trained on the JPEG restoration task. Our training data was generated by compressing high-quality images in DIV2K~\cite{div2k} and Flickr2K~\cite{flickr2k} using the OpenCV JPEG codec with uniformly sampled quality factors between 5 and 95.

At each training step, a batch of 16 compressed images was sampled, and a randomly cropped 96x96 patch was extracted from each image. An equally weighted sum of L1 and SSIM losses~\cite{ssim} was used as the training objective. We used the Adam~\cite{adam} optimizer with an initial learning rate of $2\times10^{-4}$, which was halved every $10^4$ iterations until it reached $1.25 \times 10^{-5}$. Overall, QAIRN was trained on three Tesla P-100 GPUs for 48 hours and 500,000 training steps.

\section{Evaluation}
\label{sec:evaluation}
Since QAIRN is a joint restoration and quality assessment network, we evaluated both aspects of its performance. To evaluate QAIRN's restoration performance, we sampled pristine color images from the LIVE-IQA dataset \cite{live_iqa} and compressed them at four quality factors (QFs) - 10, 20, 30, and 40. These compressed images were restored using QAIRN, and the restoration accuracy was measured using PSNR, SSIM, and PSNR-B \cite{psnrb}. A comparison of QAIRN against the state-of-the-art (SOTA) JPEG restoration models QGAC \cite{qgac} and FBCNN  is presented in Table \ref{tab:restoration}. It may be seen that despite using one network for all QFs and not using quality-aware training, QAIRN achieved near-SOTA restoration accuracy across quality factors.

The restoration results show that the use of RQAttention, which may limit the capacity of skip connections using gating, does not significantly affect restoration accuracy. We now demonstrate that the attention maps \(\gamma_n\) are indeed quality-aware. To do this, we obtained estimates of overall quality \(Q_n\) by Minkowski pooling each quality map, to obtain three measures of quality. We uses Minkowski pooling instead of simple averaging based on observations made in \cite{essim}.

\begin{table}[t]
\caption{Restoration Performance of QAIRN}
\label{tab:restoration}
\begin{center}
\begin{tabular}{|c|c|c|c|c|}
\hline
Network & QF & PSNR & SSIM & PSNR-B \\
\hline
\multirow{4}{*}{QGAC} & 10 & 27.62 & 0.804 & 27.43  \\
 & 20 & 29.88 & 0.868 & 29.56 \\
 & 30 & 31.17 & 0.896 & 30.77 \\
 & 40 & 32.05 & 0.912 & 31.61 \\
\hline
\multirow{4}{*}{FBCNN} & 10 & 27.77 & 0.803 & 27.51 \\
 & 20 & 30.11 & 0.868 & 29.70 \\
 & 30 & 31.43 & 0.897 & 30.92 \\
 & 40 & 32.34 & 0.913 & 31.80 \\
\hline
\multirow{4}{*}{QAIRN} & 10 & 27.25 & 0.803 & 26.90\\
 & 20 & 29.60 & 0.868 & 29.14\\
 & 30 & 30.94 & 0.896 & 30.41\\
 & 40 & 31.86 & 0.911 & 31.29\\
\hline
\end{tabular}
\end{center}
\end{table}

To measure quality assessment accuracy, we compute Pearson, and Spearman correlation (PCC, SRCC) between the three quality estimates and subjective opinions for JPEG distorted images from three databases - LIVE-IQA \cite{live_iqa}, CSIQ \cite{csiq}, and TID2013 \cite{tid2013}. These correlation results are presented in Table \ref{tab:jpeg_corr}. From this Table, it may be observed that \(Q_2\), derived from the second RQAttention map \(\gamma_2\), achieved a high correlation against subjective scores. 

Interestingly, both \(Q_1\) and \(Q_3\) achieved significantly lower correlations. We attribute this behavior to the tradeoff between model complexity and feature abstraction. A shallow quality map like \(\gamma_1\) is used to gate features that more closely resemble the input, making \(\gamma_1\) more interpretable as visual quality. However, the shallowness limits the capacity of RQAttention to learn strong representations. On the other hand, a deep map like \(\gamma_3\) enjoys greater model capacity but limited interpretability as visual quality, since it is used to gate deep features that are more abstract compared to the input.

\begin{table}[b]
    \centering
    \caption{Quality Assessment Performance of QAIRN}
    \label{tab:jpeg_corr}
    \begin{tabular}{|c|c|c|c|c|}
    \hline
    Database & Quality Estimate & PCC & SRCC & KCC \\
    \hline
    \multirow{3}{*}{LIVE-IQA} &  \(Q_1\), from \(\gamma_1\) & 0.099 & 0.076 & 0.051\\
     &  \(Q_2\), from \(\gamma_2\) & \textbf{0.870} & \textbf{0.879} & \textbf{0.695} \\
     &  \(Q_3\), from \(\gamma_3\) & 0.267 & 0.203 & 0.135 \\
     \hline
    \multirow{3}{*}{CSIQ} &  \(Q_1\), from \(\gamma_1\) & -0.233 & -0.216 & -0.148\\
     &  \(Q_2\), from \(\gamma_2\) & \textbf{0.859} & \textbf{0.832} & \textbf{0.628}\\
     &  \(Q_3\), from \(\gamma_3\) & 0.446 & 0.428 & 0.288\\
     \hline
    \multirow{3}{*}{TID2013} &  \(Q_1\), from \(\gamma_1\) & 0.070 & 0.118 & 0.084 \\
     &  \(Q_2\), from \(\gamma_2\) & \textbf{0.880} & \textbf{0.862} & \textbf{0.658}\\
     &  \(Q_3\), from \(\gamma_3\) & 0.073 & 0.136 & 0.078\\
     \hline
    \end{tabular}
\end{table}

Interestingly, we found that despite training only on JPEG distortions, the quality estimate \(Q_2\) displayed remarkable generalization to other distortion types, including JPEG2000 compression, Gaussian blur, and Additive White Gaussian Noise (AWGN). The correlations achieved by \(Q_2\) against human opinions on these three distortion types are presented in Table \ref{tab:other_corr}.

\begin{table}[t]
    \centering
    \caption{Generalization of QAIRN to Unseen Distortions}
    \label{tab:other_corr}
    \begin{tabular}{|c|c|c|c|c|}
    \hline
    Database & Distortion Type & PCC & SRCC & KCC \\
    \hline
    \multirow{3}{*}{LIVE-IQA} & JPEG2000 & 0.814 & 0.828 & 0.637\\
     &  Gaussian Blur & 0.714 & 0.743 & 0.557 \\
     &  Gaussian Noise & -0.895 & -0.926 & -0.778\\
     \hline
    \multirow{3}{*}{CSIQ} & JPEG2000 & 0.672 & 0.679 & 0.476\\
     &  Gaussian Blur & 0.714 & 0.767 & 0.540\\
     &  Gaussian Noise & -0.581 & -0.572 & -0.403\\
     \hline
    \multirow{3}{*}{TID2013} & JPEG2000 & 0.477 & 0.528 & 0.348\\
     &  Gaussian Blur & 0.608 & 0.654 & 0.478\\
     &  Gaussian Noise & -0.395 & -0.410 & -0.287\\
     \hline
    \end{tabular}
\end{table}

Since JPEG introduces compression distortions such as blur and blockiness, it is understandable that the metrics exhibit some generalization to JPEG2000, which is another compression method, and to Gaussian Blur. The seemingly anomalous observation is the negative correlation for Gaussian noise. In other words, the estimated quality is higher for noisier images. 

This apparent paradox may be resolved by understanding the characteristics of compressed images, which are the training inputs for QAIRN. Since compression leads to a loss of texture, input images that exhibit strong texture information are associated by the network with low compression, i.e., high quality. The addition of white noise introduces a ``grainy'' texture to images, which leads the network to assess its quality as being high. This explains the negative correlation between human subjective opinions. 

\section{Conclusion}
In this paper, we have demonstrated the use of image restoration as a proxy task to obtain an unsupervised deep quality model. We achieve this using a novel RQAttention layer that acts as a gate between input image features and processed features, depending on their quality. We showed that the use of RQAttention does not inhibit image restoration accuracy, while also demonstrating strong correlations against human subjective opinions when predicting quality.

Although QAIRN is a promising proof of concept, there is scope for improvement. Unlike FBCNN, QAIRN does not use compression quality factors as inputs during training. Hence, the same architecture can be used for a variety of restoration tasks, yielding quality metrics tuned to various distortions. In the future, we plan to generalize QAIRN to the video domain, particularly for compressed videos.

\bibliographystyle{IEEEbib}
\bibliography{test}

\end{document}